\documentclass{kluwer}    

\newdisplay{guess}{Conjecture}
\newcommand{\be}{\begin{equation}}
\newcommand{\ee}{\end{equation}}
\newcommand{\ba}{\begin{eqnarray}}
\newcommand{\ea}{\end{eqnarray}}

\newcommand{\ov}{\overline}
\newcommand{\td}{\tilde}
\newcommand{\bmatrix}[1]{\left( \begin{array}{#1}}
\newcommand{\ematrix}{\end{array}\right)}
\newcommand{\opensquare}{\mbox{$\rlap{$\sqcap$}\sqcup$}}

\newcommand{\const}{\mbox{\rm const}\,}

\begin{document}
\begin{article}
\begin{opening}
\title{Multidimensional Cosmology 
and Asymptotical AdS}
\author{U. G\"unther (1), P. Moniz (2), A. Zhuk (3)}
\runningauthor{ G\"unther, Moniz, Zhuk} \runningtitle{Multidimensional Cosmology 
and Asymptotical AdS} \institute{(1)  Inst. Math., Universit\"at Potsdam,  D-14415 Potsdam, Germany, (2) Dept. Phys., UBI, 6200
Covilh$\tilde{a}$, Portugal,
 (3) Dept. Phys, University of Odessa, Odessa 65100, Ukraine }
\date{September 5, 2002}

\begin{abstract}
A non-linear gravitational model with a
multidimensional  geometry and
 quadratic scalar curvature is considered.
For certain parameter ranges, the extra dimensions are stabilized
if the internal spaces have negative constant curvature.
As a consequence, the 4--dimensional effective cosmological constant
as well as the bulk cosmological constant become negative. The 
homogeneous and isotropic external space is asymptotically
$\mbox{AdS}_4$. The connection between the D--dimensional and the
4--dimensional fundamental mass scales sets an additional
restriction on the parameters of the considered non-linear models.
\end{abstract}

\end{opening}

\section{Introduction}
\setcounter{equation}{0}


The multidimensionality of our Universe follows naturally from
theories unifying different fundamental interactions with gravity,
e.g. M/string theory \cite{pol-wit}. This idea has received a
great deal of renewed attention over the last few years within the
"brane-world" description of the Universe. In the ADD brane model
\cite{sub-mill1}, the geometry is assumed to be factorizable as
in the standard Kaluza-Klein model, i.e.,  the topology is
the direct product of a non-warped manifold of the external
space-time and warped manifolds of the internal spaces where the
warp factors are functions of the external coordinates.

According to observations the internal space should be static or
nearly static at least from the time of primordial
nucleosynthesis (otherwise the fundamental physical constants
would vary). This means that at the present evolutionary stage of
the Universe  the compactification scale of the internal space
should either be stabilized and trapped at the minimum of some
effective potential, or it should be slowly varying (similar to
the slowly varying cosmological constant in the quintessence
scenario \cite{WCOS}). In both cases, small fluctuations over
stabilized or slowly varying compactification scales (conformal
scales/geometrical moduli) are possible.

 String theory suggests that
the usual linear Einstein-Hilbert action should be extended with
higher order non-linear curvature terms. Within this context 
we  investigate  the problem of large extra dimensions stabilization
in such theories. We find that the stabilization of extra
dimensions takes place only if additional internal spaces have a
compact hyperbolic geometry and the effective $4-$dimensional
cosmological constant is negative. Additionally, we show that
requiring  the extra dimensions to be dynamically stabilized is a
sufficient condition for the bulk space-time to acquire a constant
negative curvature. In the case that the external space $M_0$ is
homogeneous and isotropic this implies that $M_0$ becomes
asymptotically anti de Sitter (see ref. \cite{US}
 for additional details).



\section{General theory}
\setcounter{equation}{0}


We consider a $D= (4+D^{\prime})$ -- dimensional non-linear pure
gravitational theory with the action $ S = \frac
{1}{2\kappa^2_D}\int_M d^Dx \sqrt{|\ov g|} f(\ov R)\; $, where
$f(\ov R)$ is an arbitrary smooth function of a scalar
curvature $\ov R = R[\ov g]$ constructed from the D--dimensional
metric $\ov g_{AB}\; (a,b = 1,\ldots,D)$ and $  \kappa^2_D = 8\pi
/ M_{Pl(4+D^{\prime})}^{(2+D^{\prime})}$
is the D--dimensional gravitational constant. The equation of
motion for this theory reads $f^{\prime }\ov R_{ab} -\frac12 f\,
\ov g_{ab} - \ov \nabla_a \ov \nabla_b f^{\prime } + \ov g_{ab}
\ov{\opensquare } f^{\prime } = 0\;$
where $ f^{\prime } =df/d\ov R$, $\; \ov R_{ab} = R_{ab}[\ov g]$,
\ $\; \ov \nabla_a$ is the covariant derivative with respect to
the metric $\ov g_{ab}$ and the corresponding Laplacian is
denoted by  $ \ov{\opensquare}$.
For $f'(\ov R) >0$  the conformal
transformation
$ g_{ab} = \Omega^2 \ov g_{ab}\; , $
with
$ \Omega = \left[ f'(\ov R)\right]^{1/(D-2)}\;  $
reduces the non-linear theory  to a linear one with an additional
scalar field. We can further write \be G_{ab} = \phi_{,a}\phi_{,b}
-\frac12 g_{ab}g^{mn}\phi_{,m}\phi_{,n} - \frac12 g_{ab}\;
e^{\frac {-D}{\sqrt{(D-2)(D-1)}}\phi} \left(\ov R f'- f\right) \ee
and
\be \opensquare \phi = \frac {1}{\sqrt{(D-2)(D-1)}}\; e^{\frac
{-D}{\sqrt{(D-2)(D-1)}}\phi} \left(\frac {D}{2} f-f'\ov R
\right)\; , \ee
where \be f' = \frac {df}{d \ov R} := e^{A \phi} > 0\; ,\quad A :=
\sqrt{\frac{D-2}{D-1}}\; . \ee

Let us consider what will happen if, in some  way, the scalar
field $\phi$ tends asymptotically to a constant: $\phi \to
\phi_{0} \equiv \const $. As it follows, in this limit the
non-linear theory leads to a linear one: $f(\ov R) \sim c_1 \ov R
+ c_2$ with $c_1 = f' = \exp(A \phi_{0})$ and a cosmological
constant $-c_2/(2c_1)$. Thus, in the
limit $\phi \to \phi_{0}$ the D--dimensional theory
 is asymptotically (A)dS with scalar curvature: $ \ov R
\rightarrow -\frac{D}{D-2}\, \frac{c_2}{c_1}\, . $
Clearly, the linear theory  reproduces this asymptotic (A)dS-limit
for $\phi \to \phi_{0}$ with the scalar curvature:
\be R \rightarrow  2\frac{D}{D-2}\, U(\phi_0) = - \frac{D}{D-2}\;
c_2\,  c_1^{-\frac{D}{D-2}}\, . \ee
Hence, in this limit $\ov R / R \to c_1^{\frac{D}{D-2}}$  and $f'=
c_1$. We shall consider in the following the quadratic theory:
$ f(\ov R ) = \ov R + \alpha \ov R^{\; 2} - 2\Lambda _D$. 
For this theory we obtain
\be 1 + 2\alpha \ov R = e^{A \phi} \Longleftrightarrow \ov R =
\frac{1}{2\alpha } \left( e^{A \phi } - 1\right) \ee
and
\be U(\phi ) = \frac12 e^{-B \phi }\left[ \frac{1}{4\alpha }\left(
e^{A \phi } - 1\right)^2 + 2 \Lambda _D \right]\; . \ee
The condition $f' > 0$ implies $1+2\alpha \ov R
>0$.



\section{Dimensional reduction}
\setcounter{equation}{0}


To investigate the stabilization of the extra dimensions, we have
to specify the topology of the bulk space-time manifold and the
metric on it. Let the D-dimensional space-time manifold (bulk) $M$
have the form: $ M = M_0
\times M_1 \times \ldots \times M_n $ with the metric on $M$ \be
\ov g=\ov g_{ab}(X)dX^a\otimes dX^b=\ov
g^{(0)}+\sum_{i=1}^ne^{2\ov {\beta} ^i(x)}g^{(i)}\; . \ee
The coordinates on the $(D_0=d_0+1)$ - dimensional manifold $M_0 $
(usually assumed as our $(D_0=4)$-dimensional Universe) are
denoted by $x$ and the corresponding metric by
$ \ov g^{(0)}=\ov g_{\mu \nu }^{(0)}(x)dx^\mu \otimes
dx^\nu\; . $
Let the internal factor manifolds $M_i$ be $d_i$-dimensional
Einstein spaces with metric
$g^{(i)}=g^{(i)}_{m_in_i}(y_i)dy_i^{m_i}\otimes dy_i^{n_i},$ i.e.,
$ R_{m_in_i}\left[ g^{(i)}\right] =\lambda
^ig_{m_in_i}^{(i)},\qquad m_i,n_i=1,\ldots ,d_i $. For the metric
ansatz  the scalar curvature $\ov R$ depends only on $x$: $\ov
R[\ov g] = \ov R(x)$. Thus $\phi$ is also a function of $x$ :
$\phi = \phi (x)$. Without loss of generality we set the
compactification scales of the internal spaces at present time at
$\beta^i = 0 \; (i = 1,\ldots ,n)$. $D^{\prime} = D-D_0 =
\sum_{i=1}^n d_i$ is the number of the extra dimensions).


\section{Stabilization of the internal space}
\setcounter{equation}{0}


Let us consider the case of one 
internal space for simplicity\footnote{The only difference between a general model
with $n>1$ internal spaces and the particular one with $n=1$
consists in an additional diagonalization of the geometrical
moduli excitations.}. 
A conformal transformation can yield, without loss of
generality, the
corresponding action  
(see ref. \cite{US}
 for further information):
\ba S & = & \frac 1{2\kappa
_0^2}\int\limits_{M_0}d^{D_0}x\sqrt{|\td g^{(0)}|}\left\{ R\left[
\td g^{(0)}\right] - \td g^{(0) \mu \nu}
\partial_{\mu}\varphi
\partial_{\nu} \varphi - \td g^{(0) \mu \nu} \partial_{\mu}\phi
\partial_{\nu} \phi \right. \nonumber \\ & -& \left. 2U_{eff} (\varphi ,\phi ) \right\} \; , \ea
where
$ \varphi := -\sqrt{\frac{d_1(D-2)}{D_0-2}}\beta^1 $
and
\be U_{eff}(\varphi ,\phi ) = e^{2\varphi
\sqrt{\frac{d_1}{(D-2)(D_0-2)}}} \left[ -\frac12 R_1e^{2\varphi
\sqrt{\frac{D_0-2}{d_1(D-2)}}}+ U (\phi ) \right] \; . \ee

In order to obtain a stable compactification of the internal
space, the potential $U_{eff}(\varphi ,\phi )$ should have a
minimum with respect to $\varphi$ and $\phi$.
This is obvious with
respect to the field $\varphi$ because it is precisely the
stabilization of this field that we aim to achieve. It is also
clear that potential $U_{eff}(\varphi ,\phi )$ should have a
minimum with respect to $\phi$ because without stabilization of
$\phi$ the effective potential remains a dynamical function and
the extremum condition $\left. \partial U_{eff}/\partial \varphi
\right|_{\varphi = 0} =0$ is not satisfied.
Furthermore, the extrema of the potentials
$U_{eff}(\varphi ,\phi )$ and $U(\phi)$ with respect to the field
$\phi$ coincide with each other:
\be \frac{\partial U_{eff}}{\partial \phi}= e^{2\varphi
\sqrt{\frac{d_1}{(D-2)(D_0-2)}}}\quad \frac{\partial
U(\phi)}{\partial \phi}\; . \ee
 Thus, the stabilization of the
extra dimension takes place iff the field $\phi$ goes to the
minimum of the potential $U(\phi)$.
In the general case $\Lambda _D \ne 0$
it is easy to conclude that  $\left. U\right|_{min} \ge 0$ for
$\Lambda _D \ge 0$ and $\left. U\right|_{min} < 0$ for $\Lambda _D
< 0$ (in the latter case $-1<8\alpha\Lambda _D<0$). 
Moreover, the extremum and minimum existence 
conditions imply that $q=8\alpha\Lambda_D >-1$ and 
$\alpha > 0$. It can then be shown \cite{US}  that the total potential
$U_{eff}(\varphi , \phi )$ also has a global minimum in the case
when the potential $U(\phi )$ has a negative minimum. 
Furthermore, the global minimum of $U_{eff}$ is also 
negative as well as $R_1$, the scalar curvature of the 
internal space.
The condition for the stabilization of the extra dimension  thus leads
asymptotically to a negative constant curvature bulk space-time.
This  takes place for $\alpha > 0$ and
$-1<8\alpha \Lambda _D <0$. All other configurations are excluded.
Moreover, 
$\Lambda_{eff} \equiv \left.U_{eff}\right|_{min}$ plays the role
of a $D_0$ - dimensional effective cosmological constant. 

\section{Conclusion and discussion}
\setcounter{equation}{0}


We investigated a multidimensional gravitational model with
a non-Einsteinian form of the action. In particular, we assumed
that the action is an arbitrary smooth function of the scalar
curvature $f(R)$. For such models, we concentrated our attention
on the problem of extra dimension stabilization in the case of
factorizable geometry. Conformal excitations described the
internal space scale factors. A detailed stability analysis was
carried out for a model with quadratic curvature term: $f(R) = R +
\alpha R^2 -2\Lambda _D$. It was shown that a stabilization is
only possible for the parameter range $-1 < 8\alpha \Lambda _D <0
$.

This necessarily implies that the extra dimensions are stabilized
if the compact internal spaces $M_i, \ \ i=1,\ldots ,n$ have
negative constant curvatures. The 4--dimensional cosmological
constant (which corresponds to the minimum of the effective
4--dimensional potential) is also negative. As a consequence, the
homogeneous and isotropic external $(D_0=4)$-dimensional space is
asymptotically $\mbox{AdS}_{D_0}$.

{}From a cosmological perspective, it is of interest to consider the
possibility of inflation for the 4--dimensional external
space-time within  our non-linear model. For a linear
multidimensional model with an arbitrary scalar field (inflaton) 
it can be shown that there is a possibility for inflation to occur
if the scalar fields start to roll down from the region:
\be |U(\phi )| \ge |\left.U(\phi)\right|_{min}| \gg |R_1|
e^{2\varphi \sqrt{\frac{D_0-2}{d_1(D-2)}}}\; , \ee
where the effective potential reads
\be U_{eff} \approx e^{2\varphi \sqrt{\frac{d_1}{(D-2)(D_0-2)}}}
U(\phi )\; . \ee
If $ e^{\sqrt{\frac{D-2}{D-1}}\phi} \gg 1 $ 
and hence $U(\phi) \approx \frac{1}{8\alpha} e^{(2A-B)\phi} =
\frac{1}{8\alpha} \exp{\frac{D-4}{\sqrt{(D-2)(D-1)}}\phi}$, the
slow-roll parameters $\epsilon $ and $\eta _{1,2}$  read
$ \epsilon \approx \eta_1 \approx \eta_2 \approx $
$ \frac{2d_1}{(D-2)(D_0-2)} $ $ + \frac{(D-4)^2}{2(D-2)(D-1)}$.
For the dimensionality of our observable Universe $D_0=4$, these
parameters are restricted to the range
\be \frac35 \le \epsilon ,\eta_1 ,\eta_2 \le 1\quad
\mbox{for}\quad 6 \le D \le 10\; . \ee
Thus, generally speaking, the slow-roll conditions for inflation
are satisfied in this region. The scalar field $\phi$ can act as
inflaton and drive the inflation of the external space. It is
clear that the estimates point only to the possibility for inflation
to occur. For the considered model with negative effective
cosmological constant inflation is not successfully completed if
the reheating due to the decay of the $\phi -$field excitations
and gravexcitons is not sufficient for a transition to the
radiation dominated era. In any case, for scenarios with
successful transition or without, the external space has a turning
point at its maximal scale factor where the evolution changes from
expansion to contraction \cite{Felder}. 
Obviously, for such models the negative
effective cosmological constant forbids a late time acceleration
of the Universe as indicated by recent observational data. 
This problem seems however to be cured if 
our  model is generalized, e.g.,
by inclusion of additional form fields \cite{USa}.


\acknowledgements

 U.G. and A.Z. thank  UBI for kind
hospitality. The work of A.Z.
was supported by a BCC grant from CENTRA--IST  and partly
supported by the programme SCOPES  (Swiss National Science
Foundation), project No. 7SUPJ062239. U.G. acknowledges support 
from DFG grant KON/\\ 1344/2001/GU/522. Additionally, this 
work was partially supported by the grants
POCTI/32327/P/\-FIS/2000, CERN/P/FIS/43717/2001 and
CERN/P/FIS/43737/2001.


\end{article}
\end{document}